\documentclass{siamltex}
\usepackage{citesort}

\newtheorem{fact}{Fact}
\newtheorem{algorithm}{Algorithm}

\newcommand{\xx}{\pi}

\def\mod{\,\mbox{mod}\,}

\newcommand{\abk}[0]{\mbox{ABK}}

\begin{document}

\title{Reducing Randomness via Irrational Numbers\thanks{A preliminary
version appeared in {\em Proceedings of the 29th Annual {ACM}
Symposium on Theory of Computing}, pages 200--209, 1997.}}

\author{Zhi-Zhong Chen\thanks{Department of Mathematical Sciences,
Tokyo Denki University, Hatoyama, Saitama 350-03, Japan.  This work
was performed while this author was visiting Computer Science
Division, University of California, Berkeley, CA 94720.}  
\and
Ming-Yang Kao\thanks{Department of Computer Science, Yale
University. New Haven, CT 06520.  Research supported in part by NSF
Grant CCR-9531028.}}

\maketitle

\begin{abstract} We propose a general methodology for testing whether
a given polynomial with integer coefficients is identically zero.  The
methodology evaluates the polynomial at efficiently computable
approximations of suitable irrational points.  In contrast to the
classical technique of DeMillo, Lipton, Schwartz, and Zippel, this
methodology can decrease the error probability by increasing the
precision of the approximations instead of using more random bits.
Consequently, randomized algorithms that use the classical technique can
generally be improved using the new methodology.  To demonstrate the
methodology, we discuss two nontrivial applications.  The first is to
decide whether a graph has a perfect matching in parallel.  Our new NC
algorithm uses fewer random bits while doing less work than the
previously best NC algorithm by Chari, Rohatgi, and Srinivasan. 
The second application is
to test the equality of two multisets of integers.  Our new algorithm
improves upon the previously best algorithms by Blum and Kannan and can
speed up their checking algorithm for sorting programs on a large
range of inputs.
\end{abstract}



\pagestyle{myheadings} 
\markboth{\sc z.~z.~chen and m.~y.~kao}{\sc 
reducing randomness via irrational numbers}

\section{Introduction}\label{sec:intro}
Many algorithms involve checking whether certain polynomials with
integer coefficients are identically zero.  Often times, these
polynomials have exponential-sized standard representations while
having succinct nonstandard representations
\cite{Demillo78,Rowland86,Schwartz80,Zippel79}.  This paper focuses on
testing such polynomials with integer coefficients.

Given a polynomial $Q(x_1,\ldots,x_q)$ in a succinct form, a naive
method to test it is to transform it into the standard simplified form
and then test whether its coefficients are all zero.  Since $Q$ may
have exponentially many monomials, this method may take exponential
time. Let $d_Q$ be the degree of $Q$.  DeMillo and Lipton
\cite{Demillo78}, Schwartz \cite{Schwartz80} and Zippel
\cite{Zippel79} proposed an advanced method, which we call {\em the
DLSZ method}. It evaluates $Q(i_1,\ldots,i_q)$, where $i_1,\ldots,i_q$
are uniformly and independently chosen at random from a set $S$ of
$2d_Q$ integers.  This method uses $q \lceil\log (2d_Q)\rceil$ random
bits and has an error probability at most $\frac{1}{2}$.  (Every
$\log$ in this paper is to base 2.) There are three general techniques
that use additional random bits to lower the error probability to
$\frac{1}{t}$ for any integer $t > 2$.  These techniques have their
own advantages and disadvantages in terms of the running time and the
number of random bits used.  The first performs $\lceil\log t\rceil$
independent evaluations of $Q$ at $\lceil\log (2d_Q)\rceil$-bit
integers, using $q \lceil\log (2d_Q)\rceil \lceil\log t\rceil$ random
bits.  The second enlarges the cardinality of $S$ from $2d_Q$ to $t
d_Q$ and performs one evaluation of $Q$ at $\lceil\log (t
d_Q)\rceil$-bit integers, using $q\lceil\log d_Q+\log t\rceil$ random
bits.  The third is {\em probability amplification} \cite{Motwani95}.
A basic such technique works for $t \leq 2^{q\lceil\log (2d_Q)\rceil}$
by performing $t$ pairwise independent evaluations of $Q$ at
$\lceil\log (2d_Q)\rceil$-bit integers, using $2q \lceil\log
(2d_Q)\rceil$ random bits.  Stronger amplification can be obtained by
means of random walks on expanders
\cite{CohenW89,aks87,ImpagliazzoZ89}.

In \S\ref{sec:GEN}, we propose a new general methodology for testing
$Q(x_1,\ldots,x_q)$.  Our methodology computes $Q(\pi_1,\ldots,
\pi_q)$, where $\pi_1$, $\ldots$, $\pi_q$ are suitable irrational
numbers such that $Q(\pi_1,\ldots,\pi_q) = 0$ if and only if $Q(x_1,
\ldots,x_q)\equiv 0$.  Since rational arithmetic is used
in actual computers, we replace each $\pi_i$ with a rational
approximation $\pi'_i$.  A crucial question is how many bits each
$\pi'_i$ needs to ensure that $Q(\pi'_1,\ldots,\pi'_q) = 0$ if and
only if $Q(x_1,\ldots,x_q)\equiv 0$.  We give an explicit
answer to this question, from which we obtain a new randomized
algorithm for testing $Q$.  Our algorithm runs in polynomial time and
uses $\sum_{i=1}^{q}\lceil\log(d_i+1)\rceil$ random bits, where $d_i$
is the degree of $x_i$ in $Q$.  Moreover, the error probability can be
made inverse polynomially small by increasing the bit length of each
$\pi'_i$.  Thus, our methodology has two main advantages over previous
techniques:
\begin{itemize}
\item It uses fewer random bits if some $d_i$ is less than $d_Q$.
\item It can reduce the error probability without using one additional
       random bit.
\end{itemize}

In general, randomized algorithms that use the classical DLSZ method can
be improved using the new methodology.  To demonstrate the
methodology, we discuss two nontrivial applications.  In
\S\ref{sec:PMP}, the first application is to decide whether a given
graph has a perfect matching.  This problem has deterministic
polynomial-time sequential algorithms but is not known to have a
deterministic NC algorithm 
\cite{Galil82,Karp86,Micali80,Vazirani84}. We focus on solving it in
parallel using as few random bits as possible.  Our new NC algorithm
uses fewer random bits while doing less work than the previously best
NC algorithm by Chari, Rohatgi, and Srinivasan \cite{Chari95}.  In
\S\ref{sec:MET}, the second application is to test the equality of two
given multisets of integers.  This problem was initiated by Blum and
Kannan \cite{Blum95} for checking the correctness of sorting programs.
Our new algorithm improves upon the previously best algorithms
developed by them and can speed up their checking algorithm for
sorting programs on a large range of inputs.

\section{A new general methodology for testing polynomials}\label{sec:GEN}
The following notation is  used throughout this paper.
\begin{itemize}
\item
Let $Q(x_1,\ldots,x_q)$ be a polynomial with integer coefficients; 
we wish to test whether $Q(x_1,\ldots,x_q) \equiv 0$.
\item For each $x_i$, let $d_i$ be an  upper bound on
      the degree of $x_i$ in $Q$. 
      Let $k_i = \lceil\log(d_i+1)\rceil$.
\item 
Let $k = \max^q_{i=1}k_i$ and $K = \sum^q_{i=1}k_i$; $K$ is the number
of random bits used by the methodology as shown in
Theorem~\ref{th:prop}.
\item Let $d$ be an integer upper bound on the degree of $Q$; without
loss of generality, we assume $d \geq \max^q_{i=1}d_i$.
\item Let $c$ be an upper bound on the absolute value of
      a monomial's coefficient in $Q$.
\item Let $Z$ be an upper bound on the number of monomials in $Q$;
without loss of generality, we assume $Z \leq \sum^d_{i=0}q^i$.
\item 
Let $\psi = \log c+\log Z+d (\log k+\frac{\log K}{2}+\log\ln K)$.  Let
$\ell$ be an integer at least $\psi+1+\log d$; $\ell$ determines the
precision of our approximation to the irrational numbers chosen for
the variables $x_i$.
\end{itemize}
For example, if all $d_i =1$, then $k_i = 1$, $K=q$, and our goal is
to use exactly $q$ random bits, i.e., one bit per variable $x_i$.

\begin{lemma}\label{GEN:sumOfsquareroot} 
Let $p_{1,1},\ldots,p_{1,k_1},\ldots,p_{q,1},\ldots,p_{q,k_q}$ be $K$
distinct primes.  For each $p_{i,j}$, let $b_{i,j}$ be a bit.  For
each $x_i$, let $\xx_i = \sum^{k_i}_{j=1}(-1)^{b_{i,j}}\sqrt{p_{i,j}}$.
Then $Q(x_1,\ldots,x_q)$ $\not\equiv 0$ if and only if
$Q(\xx_1,\ldots,\xx_q) \not= 0$.
\end{lemma}
\begin{proof}
This lemma follows from Galois theory in algebra \cite{Morandi96}.  
Let $A_0=B_0$ be the
field of rational numbers.  For each $x_j$, let
$K_j=\sum^j_{i=1}k_i$.  Let $A_j$ be the field generated
by $\xx_1,\xx_2,\ldots,\xx_j$ over $A_0$.  Also, let $B_j$ be the field generated by
$p_{1,1},\ldots,p_{1,k_1},\ldots,p_{j,1},\ldots,p_{j,k_j}$
over $B_0$. By induction, $A_j=B_j$, the dimension of $A_j$ over $A_0$ is
$2^{K_j}$, and the dimension of $A_j$ over $A_{j-1}$ is
$2^{k_j}$.  Thus, $\xx_j$ is
not a root of any nonzero single variate polynomial over $A_{j-1}$
that has a degree less than $2^{k_j}$. 
Since $d_j < 2^{k_j}$, by induction,
$Q(\xx_1,\ldots,\xx_j,x_{j+1},\ldots,x_q) \not\equiv 0$.  The
lemma is proved at $j = q$.
\end{proof}

In light of Lemma~\ref{GEN:sumOfsquareroot}, the next algorithm tests
$Q(x_1,\ldots,x_q)$ by approximating the irrational numbers
$\sqrt{p_{i,j}}$ and randomizing the bits $b_{i,j}$.

\begin{algorithm}\label{algorithm_generic}\rm

\begin{enumerate}

\item\label{choose_bounds} 
Compute $q, d_1,\ldots,d_q, k_1,\ldots,k_q, K, d, c, Z$.

\item\label{generate_primes} Choose
$p_{1,1},\ldots,p_{1,k_1},\ldots,p_{q,1},\ldots,p_{q,k_q}$
to be the $K$ smallest primes.

\item\label{random_bits} 
Choose each $b_{i,j}$ independently with equal probability for $0$ and
$1$.

\item\label{step_l} Pick $\ell$, which determines the
precision of our approximation to $\sqrt{p_{i,j}}$.

\item\label{approximate}
For each $p_{i,j}$, compute a rational number $r_{i,j}$ from
$\sqrt{p_{i,j}}$ by cutting off the bits after the $\ell$-th bit after
the decimal point.

\item\label{evaluate}
Compute $\Delta =Q(\sum^{k_1}_{j=1}(-1)^{b_{1,j}}r_{1,j},
\ldots,\sum^{k_q}_{j=1}(-1)^{b_{q,j}}r_{q,j})$.

\item\label{output}
Output ``$Q(x_1,\ldots,x_q) \not\equiv 0$" if and only if
$\Delta\not=0$.
\end{enumerate}
\end{algorithm}

The next lemma shows how to choose an appropriate $\ell$ at Step
\ref{step_l} of Algorithm~\ref{algorithm_generic}.

\begin{lemma}\label{GEN:cor} 
If $Q(x_1,\ldots,x_q)\not\equiv 0$, then $|\Delta| \geq 2^{-\ell}$
 with probability at least $1-\frac{\psi}{\ell-1-\log d}$.
\end{lemma}
\begin{proof} 
For each combination of the bits $b_{i,j}$, $Q(\xx_1,\ldots,\xx_q)$ is
called a {\em conjugate}.  By the Prime Number Theorem \cite{LeVeque},
$\sqrt{p_{i,j}}\leq \sqrt{K}\ln K$ and thus $|\xx_i| \leq k\sqrt{K}\ln
K$.  Then, since $Q$ has at most $Z$ monomials, each conjugate's
absolute value is at most $2^{\psi} = cZ(k\sqrt{K}\ln K)^d$.  Let
$\ell'=\ell-\psi-1-\log d$.  Let $\alpha$ be the number of the conjugates
that are less than $2^{-\ell'}$. Let $\beta=2^{K}-\alpha$ be the
number of the other conjugates.  Let $\Pi$ be the product of all the
conjugates.  By Lemma~\ref{GEN:sumOfsquareroot}, $\Pi \not= 0$, and by
algebra \cite{Jacobson}, $\Pi$ is an integer. Thus, $|\Pi| \geq 1$ and
$\alpha (-\ell')+\beta \psi \geq 0$.  Hence,
$\frac{\beta}{2^K} \geq \frac{\ell'}{\ell'+\psi}$; i.e,
$|Q(\xx_1,\ldots,\xx_q))| \geq 2^{-\ell'}$ with the desired probability.
We next show that if $|Q(\xx_1,\ldots,\xx_q)| \geq 2^{-\ell'}$, then
$|\Delta| \geq 2^{-\ell}$.  Since $r_{i,j} > \sqrt{p_{i,j}} -
2^{-\ell}$, $\sum_{j=1}^{k_i}r_{i,j} > |\xx_i| - k 2^{-\ell}$.  So
approximating $p_{i,j}$ reduces each monomial term's absolute value in
$Q(\xx_1,\ldots,\xx_q)$ by at most $c(k\sqrt{K}\ln
K)^{d-1}dk2^{-\ell}$.  Thus, $|\Delta| \geq |Q(\xx_1,\ldots,\xx_q)|
- cZ(k\sqrt{K}\ln K)^d 2^{-\ell+\log d} \geq |Q(\xx_1,\ldots,\xx_q)| -
2^{-\ell'-1} \geq 2^{-\ell}$.
\end{proof}

\begin{theorem}\label{th:prop} 
For a given $t > 1$, set $\ell \geq t \psi+1+\log d$.  If
$Q(x_1,\ldots,x_m)\equiv 0$, Algorithm~\ref{algorithm_generic} always
outputs the correct answer; otherwise, it outputs the correct answer
with probability at least $1-\frac{1}{t}$.  Moreover, it uses exactly
$K$ random bits, and its error probability can be decreased by
increasing $t$ without using one additional random bit.
\end{theorem} 
\begin{proof} 
This theorem follows from Lemma~\ref{GEN:cor} immediately.
\end{proof}

Let $||Q||$ be the size of the input representation of $Q$.  The next
lemma supplements Theorem~\ref{th:prop} by discussing sufficient
conditions for Algorithm~\ref{algorithm_generic} to be efficient.

\begin{lemma}\label{GEN:final} 
With $Z = \sum^d_{i=1}q^i$, Algorithm~\ref{algorithm_generic} takes
polynomial time in $||Q||$ and $t$ under the following conditions:
\begin{itemize}
\item
The parameters $q, d_1,\ldots,d_q, d$ are at most
$(t ||Q||)^{O(1)}$ and are computable in  time polynomial in
$t ||Q||$.
\item
The parameter $c$ is at most $2^{O(t ||Q||)}$ and is computable
in  time polynomial in $t ||Q||$.
\item\label{bound_time}
Given $\ell'$-bit numbers $p'_i$, $Q(p'_1,\ldots,p'_q)$ is computable
in  time polynomial in $t ||Q||$ and $\ell'$.
\end{itemize}
\end{lemma}
\begin{proof} 
The proof is straightforward based on the following key facts. There
are at most $(t ||Q||)^{O(1)}$ primes $p_{i,j}$, which can be
efficiently found via the Prime Number Theorem.  Each $r_{i,j}$ has at
most $(t ||Q||)^{O(1)}$ bits and can be efficiently computed by,
say, Newton's method.
\end{proof}

We can scale up the rationals $r_{i,j}$ to integers and then compute
$\Delta$ modulo a reasonably small random integer. As shown in later
sections, this may considerably improve the efficiency of Algorithm
\ref{algorithm_generic} by means of the next fact.

\begin{fact}[Thrash \cite{Thrash93}]\label{prop:Thr}
Let $h \geq 3$ be an integer.  If $H$ is a subset of $\{1,
2,\ldots,h^2\}$ with $|H| \geq \frac{h^2}{2}$, then the least common
multiple of the elements in $H$ exceeds $2^h$.  Thus, for a given
positive integer $h' \leq 2^h$, a random integer from $\{1,
2,\ldots,h^2\}$ does not divide $h'$ with probability at least
$\frac{1}{2}$.
\end{fact}

\section{Application to perfect matching test}\label{sec:PMP} 
Let $G=(V,E)$ be a graph with $n$ vertices and $m$ edges.  Let $V =
\{1, 2,\ldots,n\}$.  Without loss of generality, we assume that $n$ is
even and $m \geq \frac{n}{2}$.  A {\em perfect matching} of $G$ is a
set $L$ of edges in $G$ such that no two edges in $L$ have a common
endpoint and every vertex of $G$ is incident to an edge in $L$.

Given $G$, we wish to decide whether it has a perfect matching.  This
problem is not known to have a deterministic NC algorithm.  The
algorithm of Chari , Rohatgi, and Srinivasan \cite{Chari95} uses the
fewest random bits among the previous NC algorithms.  This paper gives
a new algorithm that uses fewer random bits while doing less work.
For ease of discussion, a detailed comparison is made right after
Theorem~\ref{PMP:overall}.

\subsection{Classical ideas}\label{sec_classic}
The {\em Tutte matrix} of $G$ is the $n \times n$
skew-symmetric matrix $M$ of $m$ distinct indeterminates $y_{i,j}$:
\[
 M_{i,j} = 
          \left\{\begin{array}{rl}
                       y_{i,j} & \mbox{if $\{i, j\} \in E$ and $i < j$},\\
                      -y_{j,i} & \mbox{if $\{i, j\} \in E$ and $i > j$},\\
                       0        & \mbox{otherwise}.
                 \end{array}
          \right.
\]
Let $L = \{\{i_1, j_1\},\ldots,\{i_{\frac{n}{2}}, j_{\frac{n}{2}}\}\}$
be a perfect matching of $G$ where $i_1 < j_1, i_2 <
j_2,\ldots,i_{\frac{n}{2}} < j_{\frac{n}{2}}$ and $i_1 < i_2 < \cdots
< i_{\frac{n}{2}}$.  Let $\pi(L) = y_{i_1,j_1} y_{i_2, j_2} \cdots
y_{i_{\frac{n}{2}},j_{\frac{n}{2}}}$.  Let $\sigma(L) = 1$ or $-1$ if
the following permutation is even or odd, respectively:
\[
\left( \begin{array}{ccccc} 1 & 2 & \cdots & n-1 & n \\ i_1 & j_1 &
\cdots & i_{\frac{n}{2}} & j_{\frac{n}{2}} \end{array} \right).
\] 
Let Pf\((G) = \sum_{L} \pi(L) \sigma(L)\), where $L$ ranges
over all perfect matchings in $G$.

\begin{fact}[Fisher and Kasteleyn \cite{ber85}, Tutte \cite{Tutte52}]
\label{th:Tutte}
\begin{itemize}
\item
$\det M = \left({\rm Pf} (G)\right)^2$.
\item
$G$ has a perfect matching if and only if $\det M \not\equiv 0$.
\end{itemize}
\end{fact}

Combining Fact~\ref{th:Tutte} and the DLSZ method, Lovasz
\cite{Lovasz79} gave a randomized NC algorithm for the 
matching problem.  Since the degree of $\det M$ is at most $n$, this
algorithm assigns to each $x_{i,j}$ a random integer from
$\{1,2,\ldots,2n\}$ uniformly and independently and outputs ``$G$ has
a perfect matching" if and only if $\det M$ is nonzero at the chosen
integers. Its error probability is at most $\frac{1}{2}$, using
$m \lceil\log(2n)\rceil$ random bits. The time and processor
complexities are dominated by those of computing the determinant of an
$n \times n$ matrix with $O(\log n)$-bit integer entries.

\subsection{A new randomized NC algorithm}\label{PMP:new}
A direct application of Theorem~\ref{th:prop} to $\det M$ uses $O(m)$
random bits, but our goal is $O(n + \log m/n)$ bits.  Therefore, we
need to reduce the number of variables in $\det M$.
\begin{itemize}
\item Let $G'$ be the acyclic digraph obtained from $G$ by orienting 
      each edge $\{i ,j\}$ into the arc $(\min\{i,j\}, \max\{i,j\})$.
\item For each vertex $i$ in $G'$, let $n_i$ be the number of
      outgoing arcs from  $i$. 
\item Let $\hat{n}_i = 0$ if $n_i = 0$;
      otherwise, $\hat{n}_i=\lceil\log n_i\rceil$.
\item Let $q = \sum^n_{i=1}\hat{n}_i$.
Note that $q < n+n\log\frac{m}{n}$.
\item Let $x_1$, $x_2$, $\ldots$, $x_q$ be $q$ distinct new indeterminates.
\end{itemize}

We label the outgoing arcs of each vertex as follows.  If $n_1 = 0$,
then vertex 1 has no outgoing arc in $G'$.  If $n_1 = 1$, then label its unique
outgoing arc with 1.  If $n_1 \geq 2$, then label its $n_1$ outgoing arcs
each with a distinct monomial in $\{(x_1)^{a_1}(x_2)^{a_2}\cdots
(x_{\hat{n}_1})^{a_{\hat{n}_1}}~|~\mbox{each $a_h$ is 0 or 1}\}$,
which is always possible since $2^{\hat{n}_1}\geq n_1$.  We label
the $n_2$ outgoing arcs of vertex 2 in the same manner using
$x_{\hat{n}_1+1}, x_{\hat{n}_1+2},\ldots,
x_{\hat{n}_1+\hat{n}_2}$.  We similarly process the other vertices
$i$, each using the next $\hat{n}_i$ avaliable indeterminates $x_h$.

Let $f_{i,j}$ be the label of arc $(i,j)$ in $G'$.  Let
$Q(x_1,\ldots,x_q)$ be the polynomial obtained from ${\rm Pf}(G)$ by
replacing each indeterminate $y_{i,j}$ with $f_{i,j}$.
\begin{lemma}\label{PMP:match} 
$G$ has a perfect matching if and only if $Q(x_1,\ldots,x_q)\not\equiv
0$.
\end{lemma}
\begin{proof} 
For each $L$ as described in \S\ref{sec_classic}, let $Q_L = \sigma(L)
f_{i_1,j_1} f_{i_2,j_2} \cdots f_{i_{\frac{n}{2}},
j_{\frac{n}{2}}}$.  Then, $Q = \sum_L Q_L$, where $L$ ranges over all
the perfect matchings of $G$.  It suffices to prove that for distinct
perfect matchings $L_1$ and $L_2$, the monomials $Q_{L_1}$ and
$Q_{L_2}$ differ by at least one $x_h$.  Let $H$ be the subgraph of
$G$ induced by $(L_1\cup L_2) - (L_1 \cap L_2)$.  $H$ is a set of
vertex-disjoint cycles.  Since $L_1 \neq L_2$, $H$ contains at least
one cycle $C$.  Let $C'$ be the acyclic digraph
obtained from $C$ by replacing each edge $\{i,j\}$ with the arc
$(\min\{i,j\},\max\{i,j\})$.  $C'$ contains two outgoing arcs
$(i,j_1)$ and $(i,j_2)$ of some vertex $i$. So there is an
indeterminate $x_h$ used in arc labels for vertex $i$, whose degree is
1 in one of $f_{i, j_1}$ and $f_{i, j_2}$ but is 0 in the other.
Hence, the degree of $x_h$ is 1 in one of $Q_{L_1}$ and $Q_{L_2}$ but
is 0 in the other, which makes $Q_{L_1}$ and $Q_{L_2}$ distinct as
desired.
\end{proof}

To test whether $G$ has a perfect matching, we use
Algorithm~\ref{algorithm_generic} to test $Q$ by means of
Theorem~\ref{th:prop} and Lemma~\ref{PMP:match}.  Below we detail each
step of Algorithm~\ref{algorithm_generic}.

Step~\ref{choose_bounds}.  Compute $q$. Then set
$d_1=d_2=\cdots=d_q=1$, $k_1=k_2=\cdots=k_q=1$, $K=q$, $d=q$, $c=1$.
Further set $Z=(\frac{2m}{n})^n$ since the number of perfect matchings
in $G$ is at most $\Pi_{i=1}^{n}m_i \leq (\frac{2m}{n})^n$, where
$m_i$ is the degree of node $i$ in $G$.

Step~\ref{generate_primes}.  This step computes the $q$ smallest
primes $p_{1,1}$, $p_{2,1}$, \ldots, $p_{q,1}$, each at most $q \ln^2q$.
Since a positive integer $p$ is prime if and only if it is indivisible
by any integer $i$ with $2 \leq i \leq \sqrt{p}$, these primes can be
found in $O(\log q)$ parallel arithmetic steps on integers of at most
$\lceil\log (1+q\ln^2q)\rceil$ bits using
$O(q^{1.5}\log^3q)$ processors.

Step \ref{random_bits}. This step is straightforward.

Step~\ref{step_l}.  Set $\ell= \lceil{t \psi}\rceil+\lceil{q}\rceil+1$, where
$\psi= n\log\frac{2m}{n}+q\log(\sqrt{q}\ln q)$.

Step~\ref{approximate}.  We use Newton's method to compute $r_{i,1}$
from $p_{i,1}$.  For the convenience of the reader, we briefly sketch
the method here.  We use $g_0 = p_{i,1}$ as the initial estimate.
After the $j$-th estimate $g_j$ is obtained, we compute
$g_{j+1}=\frac{1}{2}(g_j+\frac{p_{i,1}}{g_j})$, maintaining only the
bits of $g_{j+1}$ before the $(\ell+1)$-th bit after the decimal
point.  Thus, $g_{j+1} \leq \frac{1}{2}(g_j+\frac{p_{i,1}}{g_j})$.
With $g_{j+1}$ obtained, we check whether $g^2_{j+1} > p_{i,1}$.  If
not, we stop; otherwise, we proceed to compute $g_{j+2}$.  Since the
convergence order of the method is 2, we take the
$\lceil\log(\lceil\log p_{i,1}\rceil+\ell)\rceil$-th estimate as
$r_{i,1}$.  So $r_{1,1},\ldots,r_{q,1}$ can be computed in
$O\left(\log(\ell+\log q)\right)$ parallel arithmetic steps with $q$
processors.  Note that each $g_j$ has at most
$\lceil\log(1+q\ln^2q)\rceil+\ell$ bits.

Step~\ref{evaluate}.  Evaluating $\Delta$ is equivalent to
computing $\Delta^2$.  $\Delta^2$ is the determinant of an $n \times n$
skew-symmetric matrix $M'$ whose nonzero entries above the main
diagonal in the $i$-th row are either 1 or products of at most
$\hat{n}_i$ rationals among $r_{1,1},\ldots,r_{q,1}$.  Thus, each
matrix entry has at most
$\lceil\log{n}\rceil(\lceil\log(1+q\ln^2q)\rceil+\ell)$ bits.  Setting
up $M'$ takes $O(\log n)$ arithemetic steps on $O(n^2)$ processors.

Step \ref{output}. This step is straightforward.

The next theorem summarizes the above discussion. 

\begin{theorem}\label{PMP:overall}
For any given $t>1$, whether $G$ has a perfect matching can be
determined in $O(\log (nt))$ parallel arithmetic steps on rationals of
$O(tn\log^3n)$ bits using $O(n^2)$ processors together with one
evaluation of the determinant of an $n\times n$ matrix of
$O(tn\log^3n)$-bit rational entries.  The error probability is at most
$\frac{1}{t}$, using $q < n+n\log\frac{m}{n}$ random bits.
\end{theorem}

{\it Remark.} The best known NC algorithm for computing the
determinant of an $n \times n$ matrix takes $O(\log^2 n)$ parallel
arithmetic steps using $O(n^{2.376})$ processors \cite{Pan87}.

\begin{proof}
We separate the total complexity of Algorithm~\ref{algorithm_generic}
into that for computing $\det M'$ and that for all the other
computation. For the latter, the running time is dominated by that of
Step~\ref{approximate}; the bit length by that of the entries in $M'$
at Step~\ref{evaluate}; and the processor count by that of setting up
$M'$.
\end{proof}

The work of Chari , Rohatgi, and Srinivasan \cite{Chari95} aims to use
few random bits when the number of perfect matchings is small.
Indeed, their algorithm uses the fewest random bits among the previous
NC algorithms.  For an error probability at most $\frac{3}{4}$, it
uses $\min\{28\sum^n_{i=1}\lceil\log \hat{d}_i\rceil, \, 6m
+4\sum^n_{i=1}\lceil\log \hat{d}_i\rceil\} + O(\log n)$ random bits,
where $\hat{d}_i$ is the degree of vertex $i$ in $G$.  It also
computes the determinant of an $n \times n$ matrix with $O(n^7)$-bit
entries.  In contrast, with $t=2$ in Theorem~\ref{PMP:overall},
Algorithm~\ref{algorithm_generic} has an error probability at most
$\frac{1}{2}$ while using fewer random bits, i.e., $q <
n+n\log\frac{m}{n}$ bits. Moreover, using the best known NC algorithm
for determinants, the work of Algorithm~\ref{algorithm_generic} is
dominated by that of computing the determinant of an $n \times n$
matrix with entries of shorter length, i.e, $O(n\log^3n)$ bits.

The next theorem modifies the above implementation of
Algorithm~\ref{algorithm_generic} by means of Fact~\ref{prop:Thr} so
that it computes the determinants of matrices with only $O(\log
(nt))$-bit integer entries but uses slightly more random bits.

\begin{theorem}\label{PMP:overall2}  
For any given $t>2$, whether $G$ has a perfect matching can be
determined in $O(\log (nt))$ parallel arithmetic steps on rationals of
$O(tn\log^3n)$ bits using $O(n^2)$ processors together with
$\lceil\log t\rceil$ evaluations of the determinant of an $n\times n$
matrix of $O(\log(nt))$-bit integer entries.  The error probability is
at most $\frac{2}{t}$, using $q+O(\log t \log(nt))$ random bits, which
is at most $n+n\log\frac{m}{n}+O(\log t \log(nt))$.
\end{theorem}
\begin{proof}
We modify Steps~\ref{evaluate} and \ref{output} of the above
implementation as follows.

Step~\ref{evaluate}. 
\begin{itemize}
\item 
Compute $M'$ as above.
\item 
For each $(i,j)$-th entry of $M'$, we multiply it with
$2^{(\hat{n}_i+\hat{n}_j)\ell}$ in $O(1)$ parallel arithmetic
steps using $O(n^2)$ processors.  Let $M''$ be the resulting matrix;
note that $\det{M''}=2^{2q\ell}\det{M'}$ and each entry of $M''$ is an
integer of at most $3\lceil\log n\rceil(\ell+\lceil\log n\rceil)$
bits.
\item 
Let $\lambda = \lceil \log t \rceil$.  Let $u =
n!{\cdot}2^{3n\lceil\log n\rceil(\ell+\lceil\log n\rceil)}$; note that
$|\det M''| \leq u$.  We uniformly and independently choose $\lambda$
random positive integers $w \leq \lceil\log u\rceil^2$ using
$O(\lambda\log(nt))$ random bits in $O(\lambda)$ steps on a single
processor.  For each chosen $w$, we first compute $M''' = M'' \mod w$
in $O(1)$ parallel arithmetic steps using $O(n^2)$ processors; and
then compute $\det M'''$ instead of $\det M'$.
\end{itemize}

Step~\ref{output}. Output ``$G$ has a perfect matching" if and only if
some $\det M'''$ is nonzero.

By Fact~\ref{prop:Thr}, if $\det M'' \not= 0$, then some chosen $w$
does not divide $\det M''$ with probability at least $1-2^{-\lambda}$.
Thus, the overall error probability is at most
$\frac{1}{t}+2^{-\lambda} \leq \frac{2}{t}$.  We separate the total
complexity of Algorithm~\ref{algorithm_generic} into that for
computing $\det M'''$ and that for all the other computation.  As with
Theorem~\ref{PMP:overall}, the running time of the latter remains
dominated by that of Step~\ref{approximate}; the bit length by that of
the entries in $M'$ at Step~\ref{evaluate}; and the processor count by
that of setting up $M'$.
\end{proof}

\section{Application to multiset equality test}\label{sec:MET} Let
$A=\{a_1,\ldots,a_n\}$ and $B=\{b_1,\ldots,b_n\}$ be two multisets of
positive integers.  Let $a$ be the largest possible value for any
element of $A \cup B$. Given $A, B$, and $a$ as input, the {\em
multiset equality test problem} is that of deciding whether $A \equiv
B$, i.e., whether they contain the same number of copies for each
element in $A \cup B$.  This problem was initiated by Blum and Kannan
\cite{Blum95} to study how to check the correctness of sorting
programs.  They gave two randomized algorithms on a useful model of
computation which reflects many sorting scenarios better than the
usual RAM model.  For brevity, we denote their model by MBK and the
two algorithms by $\abk_1$ and $\abk_2$.

This section modifies the MBK model to cover a broader range of
sorting applications. It then gives a new randomized algorithm, which
improves upon $\abk_1$ and $\abk_2$ and can speed up the checking
algorithm for sorting by Blum and Kannan \cite{Blum95} on a large
range of inputs.

\subsection{Models of computation and previous results}\label{MET:prev}
In both the MBK model and the modified model, the computer has $O(1)$
tapes as well as a random access memory of $O(\log n+\log a)$ words.
The allowed elementary operations are $+$, $-$, $\times$, $/$, $<$,
$=$, and two bit operations shift-to-left and shift-to-right, where
$/$ is integer division.  Each of these operations takes one step on
integers that are one word long; thus the division of an integer of
$m_1$ words by another of $m_2$ words takes $O(m_1m_2)$ time.  In
addition, it takes one step to copy a word on tape to a word in the
random access memory or vice versa.

The only difference between the two models is that the modified model
has a shorter word length relative to $a$ and therefore is applicable
to sorting applications with a larger range of keys.  To be precise,
in the MBK model, each word has $1+\lfloor \log a \rfloor$ bits, and
thus can hold a nonnegative integer at most $a$.  In the modified
model, each word has $\xi =
1+\lfloor\log\max\{\lceil\log{n}\rceil,\lceil\log{a}\rceil\}\rfloor$
bits, and thus can hold a nonnegative integer at most
$\max\{\lceil\log{n}\rceil,\lceil\log{a}\rceil\}$.

Note that sorting $A$ and $B$ by comparison takes $O(n\log n)$ time in
the MBK model and $O(\frac{\log{a}}{\xi}n\log{n})$ time in the
modified model.  However, in both models, if $n \geq 2^a$, the
equality of $A$ and $B$ can be tested in optimal $O(n)$ time with
bucket sort.  Hence, we hereafter assume $n < 2^a$.  We briefly review
$\abk_1$ and $\abk_2$ as follows.

Let $Q_1(x)$ be the polynomial $\sum_{i=1}^n x^{a_i} - \sum_{i=1}^n
x^{b_i}$.  $\abk_1$ selects a random prime $w \leq 3a\lceil\log
(n+1)\rceil$ uniformly and computes $Q_1(n+1) \mod w$ in a
straightforward manner.  It outputs ``$A \equiv B$'' if and only if
$Q_1(n+1) \mod w$ is zero.  Excluding the cost of computing $w$,
$\abk_1$ takes $O(n\log a)$ time in the MBK model and
$O\left((\frac{\log{a}}{\xi})^2n\log{a}\right)$ time in the
modified model.  The error probability is at most $\frac{1}{2}$.

Let $Q_2(x)$ be the polynomial
$\Pi^n_{i=1}(x-a_i)-\Pi^n_{i=1}(x-b_i)$.  $\abk_2$ uniformly selects a
random positive integer $z \leq 4n$ and a random prime $w
\leq 3n\lceil\log(a+4n)\rceil$; and  computes $P(z) \mod w$ in a
straightforward manner.  It outputs ``$A \equiv B$'' if and only if
$P(z) \mod w$ is zero. Excluding the cost of computing $w$, $\abk_2$
takes $O\left(n\max\{1,(\frac{\log n}{\log a})^2\}\right)$ time in the
MBK model and
$O\left(n\frac{(\log{n}+\log{a})(\log{n}+\log\log{a})}{\xi^2}\right)$
time in the modified model.  The error probablity is at most
$\frac{3}{4}$.

Generating the random primes $w$ is a crucial step of $\abk_1$ and
$\abk_2$. It is unclear how this step can be performed efficiently in
terms of running time and random bits.  We modify this step by means
of Fact~\ref{prop:Thr} as follows.  In $\abk_1$, $|Q_1(n+1)| \leq
2^{1+a\log(n+1)+\log n}$; in $\abk_2$, $|Q_2(2n)| \leq
2^{1+n\log(a+4n)}$.  Thus, we can replace $w$ in $\abk_1$ and $\abk_2$
with two random positive integers $w_1 \leq (1+a\log(n+1)+\log n)^2$
and $w_2 \leq (1+n\log(a+4n))^2$, respectively.  With these
modifications, $\abk_1$ and $\abk_2$ use at most $2\log a + 2\log\log
n +O(1)$ and $3\log n + 2\log\log(a+n) + O(1)$ random bits,
respectively. The time complexities and error probabilties remain as
stated above.

\subsection{A new randomized algorithm}\label{MET:newalgo}
Our goal in this section is to design an algorithm for multiset
equality test for the modified model that is faster than $\abk_1$ for
$n=\omega((\log\log a)^2)$ and faster than $\abk_2$ for
$n=\omega\left((\log{a})^{\log\log{a}}\right)$.  We can then use it to
speed up the previously best checking algorithm for sorting
\cite{Blum95}.

\begin{itemize}
\item Let $q = \lfloor\log a\rfloor +1$. 
\item Let $x_1$, $\ldots$, $x_q$ be $q$ distinct indeterminates.
\item For each $u\in A\cup B$, let $f_u$ denote the monomial
$(x_1)^{u_1}(x_2)^{u_2}\cdots(x_q)^{u_q}$, where $u_1u_2\cdots u_q$ is
the standard $q$-bit binary representation of $u$.
\item 
Let $Q(x_1,\ldots,x_q)$ denote the polynomial
$\sum^{n}_{i=1}f_{a_i}-\sum^{n}_{i=1}f_{b_i}$.
\end{itemize}
Note that $Q(x_1,\ldots,x_q) \equiv 0$ if and only if $A \equiv B$.
To test whether $A \equiv B$, we detail how to implement the steps of
Algorithm~\ref{algorithm_generic} to test $Q$ as follows.  
The algorithm is analyzed only with respect to the modified model.

{\it Remark.} In the implementation, the parameter $t$ of
Theorem~\ref{th:prop} needs to be a constant so that the algorithm can
be performed inside the random access memory together with
straightforward management of the tapes. At the end of this section,
we set $t =4$ but for the benefit of future research, we analyze the
running time and the random bit count in terms of a general $t$.

Step~\ref{choose_bounds}.  Compute $q$ by finding the index of the
most significant bit in the binary representation of $a$. Since $a$
takes up $O(\frac{\log a}{\xi})$ words, this computation takes
$O(q)$ time by shifting the most significant nonzero word
to the left at most $\xi$ times.  Afterwards, set
$d_1=d_2=\cdots=d_q=k_1=k_2=\cdots=k_q=k=1$, $K=d=q$, $c=n$, and $Z=2n$
in $O(q)$ time. This step takes $O(q)$ time.

Step~\ref{generate_primes}.  Compute the $q$ smallest primes $p_{1,1},
p_{2,1}, \ldots, p_{q,1} \leq q\ln^2q$.  We compute these primes by
inspecting $i=2, 3, \ldots$ one at a time up to $q\ln^2q$ until
exactly $q$ primes are found. Since $i$ can fit into $O(1)$ words, it
takes $O(\sqrt{q}\log q)$ time to check the primality of each $i$
using the square root test for primes in a straightforward manner.
Thus, this step takes $O(q^{3/2}\log^3q)$ time.

Step \ref{random_bits}. This step is straightforward and uses $q$
random bits and $O(\frac{q}{\xi})$ time.

Step~\ref{step_l}.  Set $\ell= \lceil t \rceil \psi' +\lceil q \rceil+1$, 
where $t$
is a given positive number and $\psi'= 2\lceil \log n \rceil +
\lceil\frac{q\lceil\log q\rceil}{2}\rceil +q\lceil\log\lceil\log
q\rceil\rceil+1$.  The number $\lceil\log n\rceil$ can be computed
from the input in $O(n)$ time. The computations of $\lceil\frac{\log
q}{2} \rceil$ and $\lceil\log\lceil\log q \rceil\rceil$ are similar to
Step~\ref{choose_bounds} and take $O(\log q)$ time.  Thus, this step
takes $O(n+\log q+\frac{\log t}{\xi})$ time.

Step~\ref{approximate}.  As at Step~\ref{approximate} in
\S\ref{PMP:new}, we use Newton's method to compute $r_{i,1}$ for each
$p_{i,1}$.  With only integer operations allowed, we use $2^{\ell}g_j$
as the $j$-th estimate for $2^{\ell}\sqrt{p_{i,1}}$; i.e.,
$2^{\ell}g_{j+1}=(2^{\ell}g_j+2^{2\ell}p_{i,1}/(2^{\ell}g_j))/2$.  The
last estimate computed in this manner is $2^{\ell}r_{i,1}$. Since
$2^\ell$ can be computed in $O((\frac{\ell}{\xi})^2)$ time using a
doubling process, the first estimate $2^{\ell}p_{i,1}$ can be computed
in the same amount of time. Since the other estimates all are
$O(\frac{\ell}{\xi})$ words long, the $(j+1)$-th estimate can be
obtained from the $j$-th in $O((\frac{\ell}{\xi})^2)$ time. Since
only $O(\log \ell)$ iterations for each $2^{\ell}\sqrt{p_{i,1}}$ are
needed, this step takes $O(q(\frac{\ell}{\xi})^2\log\ell)$ time.

Step~\ref{evaluate}.  We compute
$\Delta=Q((-1)^{b_{1,1}}r_{1,1},\ldots,(-1)^{b_{q,1}}r_{q,1})$ by
means of Fact~\ref{prop:Thr} as follows.  Let $\lambda = \lceil \log t
\rceil$.  Since $|2^{q\ell}\Delta|$ is an integer at most
$2^{\psi'+q\ell}$, we uniformly and independently select $\lambda$
random positive integers $w \leq (\psi'+q\ell)^2$ using $2\lambda
(\log t + \log\log n +2\log\log a + o(\log\log a))$ random bits and
$O(\lambda \frac{\log \ell}{\xi})$ time.  Note that if
$2^{q\ell}\Delta\not=0$, then with probability at least
$1-\frac{1}{t}$, some $2^{q\ell}\Delta\mod{w}$ is nonzero.  We next
compute all $2^{ql}\Delta\mod{w}$.
For each element $u\in A\cup B$, let $e(u)$ be
the number of 0's in the standard $q$-bit binary representation of
$u$.  Let
$h(u)=f_u((-1)^{b_{1,1}}2^{\ell}r_{1,1},\ldots,(-1)^{b_{q,1}}2^{\ell}r_{q,1})$.
Then, $2^{q\ell}\Delta=\sum_{i=1}^n2^{e(a_i)\ell}h(a_i) -
\sum_{i=1}^n2^{e(b_i)\ell}h(b_i)$, which we use to compute
all $2^{q\ell}\Delta\mod{w}$ as follows.
\begin{itemize}
\item
Compute the numbers $e(u)$ for all $u\in A \cup B$ in $O(nq)$ time.
\item 
For all $w$, compute all $2^{\ell}r_{i,1}\mod w$ in
$O(\lambda q\frac{\ell}{\xi}\frac{\log\ell}{\xi})$ time.
\item 
For all $w$, use values obtained above to compute $h(u)\mod w$ for all
$u$ in $O(\lambda n q (\frac{\log\ell}{\xi})^2)$ time.
\item  
For all $w$, compute $2^{\ell} \mod w$ in $O(\lambda
\frac{\ell}{\xi}\frac{\log\ell}{\xi})$ time.
\item 
For all $w$, use values obtained above to compute $2^{e(u)\ell} \mod
w$ for all $u$ in $O(\lambda n(\frac{\log\ell}{\xi})^2\log q)$
time.
\item 
For all $w$, use values obtained above to compute $2^{q\ell}\Delta
\mod w$ in $O(\lambda n (\frac{\log\ell}{\xi})^2)$ time.
\end{itemize}
This step uses $2\lambda (\log t + \log\log n +2\log\log a +
o(\log\log a))$ random bits and takes $O(\lambda
q\frac{\ell}{\xi}\frac{\log\ell}{\xi} +
\lambda n q (\frac{\log\ell}{\xi})^2)$ time.

Step \ref{output}. Output ``$A \not\equiv B$'' if and only if some
$2^{ql}\Delta \mod w$ is nonzero.

The next theorem summarizes the above discussion.

\begin{theorem}\label{MET:overall} 
For any given $t>2$, whether $A\equiv B$ can be determined in time
\[O\left(q\log\ell\left(\frac{\ell}{\xi}\right)^2+ 
\lambda n q \left(\frac{\log \ell}{\xi}\right)^2\right),\] where $q =
\Theta(\log a); \ell = \Theta(t(\log n + \log a \log\log a)); \xi =
\Theta(\log\log(n+a)); \lambda = \Theta(\log t).$ The error
probability is at most $\frac{2}{t}$ using $\log a + 2 \lceil \log t
\rceil (\log t + \log\log n + 2 \log\log a + o(\log\log a))$ random
bits.
\end{theorem}
\begin{proof}
The running time of Algorithm~\ref{algorithm_generic} is dominated by
those of Steps~\ref{approximate} and \ref{evaluate}.  The error
probability follows from Theorem~\ref{th:prop} and
Fact~\ref{prop:Thr}.
\end{proof}

We use the next corollary of Theorem~\ref{MET:overall} to compare
Algorithm~\ref{algorithm_generic} with $\abk_1$ and $\abk_2$ in the
modified model.
\begin{corollary}\label{last_cor}
With $t=4$, Algorithm~\ref{algorithm_generic} has an error probability
at most $\frac{1}{2}$ using $\log a + 4\log\log n + 8 \log\log a +
o(\log\log a)$ random bits, while running in time
\[
O\left(n \log a + \log a \frac{(\log n + \log a \log\log
a)^2}{\log\log(n+a)}\right).
\]
\end{corollary}

By corollary \ref{last_cor}, Algorithm~\ref{algorithm_generic} is 
faster than $\abk_1$ for $n=\omega((\log\log a)^2)$ and faster than
$\abk_2$ for $n = \omega\left((\log{a})^{\log\log{a}}\right)$.  Thus,
it can replace $\abk_1$ and $\abk_2$ to speed up the previously best
checking algorithm for sorting \cite{Blum95} as follows. We use bucket
sort for $2^a \leq n$; Algorithm~\ref{algorithm_generic} for
$(\log{a})^{\log\log{a}} \leq n < 2^a$; and $\abk_2$ otherwise.

\section*{Acknowledgments}
We are very grateful to Steve Tate for useful discussions and to the
anonymous referees for extremely thorough and helpful comments.

%\bibliographystyle{siam} 
%\bibliography{all}

\end{document}